# Detection and Severity Classification of COVID-19 in CT images using deep learning

Yazan Qiblawey[a], Anas Tahir[a], Muhammad E. H. Chowdhury[a*], Amith Khandakar[a], Serkan Kiranyaz[a], Tawsifur Rahman[b], Nabil Ibtehaz[c], Sakib Mahmud[a], Somaya Al-Madeed[d], Farayi Musharavati[e]

[a]Department of Electrical Engineering, Qatar University, Doha-2713, Qatar; yazan.qiblawey@qu.edu.qa (YQ), a.tahir@qu.edu.qa (AT), mchowdhury@qu.edu.qa (MEHC), amitk@qu.edu.qa (AK), mkiranyaz@qu.edu.qa (SK), sm1512633@qu.edu.qa (SM)
[b]Department of Biomedical Physics & Technology, University of Dhaka, Dhaka-1000, Bangladesh; tawsifurrahman@bmpt.du.ac.bd (TR)
[c]Department of Computer Science and Engineering, Bangladesh University of Engineering and Technology, Dhaka-1205, Bangladesh; 1017052037@grad.cse.buet.ac.bd
[d]Department of Computer Science and Engineering, Qatar University, Doha-2713, Qatar; s_alali@qu.edu.qa
[e]Mechanical & Industrial Engineering Department, Qatar University, Doha-2713, Qatar farayi@qu.edu.qa

*Corresponding author: Muhammad E. H. Chowdhury (mchowdhury@qu.edu.qa)

**Abstract**

Since the breakout of coronavirus disease (COVID-19), the computer-aided diagnosis has become a necessity to prevent the spread of the virus. Detecting COVID-19 at an early stage is essential to reduce the mortality risk of the patients. In this study, a cascaded system is proposed to segment the lung, detect, localize, and quantify COVID-19 infections from computed tomography (CT) images Furthermore, the system classifies the severity of COVID-19 as mild, moderate, severe, or critical based on the percentage of infected lungs. An extensive set of experiments were performed using state-of-the-art deep Encoder-Decoder Convolutional Neural Networks (ED-CNNs), UNet, and Feature Pyramid Network (FPN), with different backbone (encoder) structures using the variants of DenseNet and ResNet. The conducted experiments showed the best performance for lung region segmentation with Dice Similarity Coefficient (DSC) of 97.19% and Intersection over Union (IoU) of 95.10% using U-Net model with the DenseNet 161 encoder. Furthermore, the proposed system achieved an elegant performance for COVID-19 infection segmentation with a DSC of 94.13% and IoU of 91.85% using the FPN model with the DenseNet201 encoder. The achieved performance is significantly superior to previous methods for COVID-19 lesion localization. Besides, the proposed system can reliably localize infection of various shapes and sizes, especially small infection regions, which are rarely considered in recent studies. Moreover, the proposed system achieved high COVID-19 detection performance with 99.64% sensitivity and 98.72% specificity. Finally, the system was able to discriminate between different severity levels of COVID-19 infection over a dataset of 1,110 subjects with sensitivity values of 98.3%, 71.2%, 77.8%, and 100% for mild, moderate, severe, and critical infections, respectively.

Keywords: COVID-19, Lung Segmentation, Lesion Segmentation, Severity Classification, Deep Learning

# 1   Introduction

The coronavirus disease 2019 (COVID-19) has become a global pandemic, which affects different aspects of human life. Until the 11 of January 2020, more than 88.8 million confirmed cases and 1.92 million death cases have been recorded and its infection rate is still rapidly increasing worldwide [1]. Several laboratory identification tools are used for the detection of COVID-19, such as real-time reverse transcription-polymerase chain reaction (RT-PCR) and isothermal nucleic acid amplification technology [2, 3]. Currently, RT-PCR is considered the gold standard to detect COVID-19 [4]. However, a high false alarm rate usually occurs due to the sample contamination, damage, or virus mutations in the COVID-19 genome. Medical imaging can be considered as a first-line investigation tool [5]. Several studies [6, 7] suggested performing chest computerized tomography (CT) image as a secondary test if the suspected patients show symptoms after a negative RT-PCR finding. For instance, in Wuhan, China, among 1014 COVID-19 patients, 59% had positive RT-PCR results but 88% had positive CT scans. Besides, among the positive RT-PCR results, the CT scans achieved a 97% sensitivity [8]. Thus, CT scans can detect COVID-19 with higher accuracy than RT-PCR. Moreover, CT images can show early lesions in the lung and they can be used for the diagnosis by radiologists. However, radiologists require significant diagnostic experience to distinguish COVID-19 from other types of pneumonia [9]. Radiologists need to carry out two tasks for COVID-19 patients which are identification and severity quantification. The purpose of identification is to identify COVID-19 patients among other patients to isolate them as early as possible. Severity quantification can help medical personnel to prioritize the patients who will require emergency medical care. It requires a high evaluation time for radiologists to carry out both tasks. Thus, developing artificial intelligence (AI)-based solutions specific to identification and severity quantification to COVID-19 can offer a fast, efficient and reliable alternative that can supplement conventional medical diagnostic strategies. Recent studies showed that state-of-the-art deep convolutional neural networks (CNNs) can achieve or exceed the performance of medical experts in numerous medical image diagnosis tasks, such as skin lesion classification [10], brain tumor detection [11], and breast cancer detection [12], and lung pathology screening [13, 14].

*1.1. Related Work*

In general, COVID-19 recognition from other types of pneumonia has a unique difficulty compared to other lung diseases, such as tuberculosis screening, lung nodule detection, and lung cancer diagnosis. This difficulty arises from the high similarity between different types of pneumonia (especially in the early-stage) and large variations in various stages of the same type. Powered by large annotated datasets and modern graphical processing units (GPUs), machine learning especially deep learning techniques, has achieved outbreak performance in several computer vision applications, such as image classification, object detection, and image segmentation. Recently deep learning techniques on chest CT scans and chest X-ray (CXR) images are getting increased popularity for diagnosing different lung diseases, showing promising results in various applications. Several studies have been published on CT-based COVID-19 diagnosis systems using machine learning models [15-19]. Several representative studies are summarised and reviewed below. Harmon et al. [20] trained and evaluated a series of deep learning networks on a diverse multi-national cohort of 922 COVID-19 cases and 1695 non-COVID patients to localize lung parenchyma followed by identification of COVID-19 pneumonia. AH-Net was utilized for lung volume segmentation achieving a dice similarity coefficient (DSC) of 95%, while 3D-Densnet-121 was employed to recognize lung regions as COVID-19 or non-COVID. The average score of multiple lung regions was utilized for the classification scheme achieving 88.9% accuracy, 85.3% sensitivity, and 90.1% specificity. Wang et al. [21] introduced a deep regression framework for automatic pneumonia identification by jointly learning from CT scan images and clinical information (i.e., age, gender, and clinical complaints). Recurrent Neural Network (RNN) with ResNet50 as the backbone was used to extract visual features from CT images. The initial clinical information collected from admitted patients (fever, cough, trouble in breathing, etc.) was analyzed by a Long short-term memory (LSTM) network and concatenated with demographic features (age and gender), and extracted visual features from CT images. Finally, a regression framework was utilized to diagnose the suspected patient as Community-acquired pneumonia (CAP) or normal. The proposed framework was evaluated over 900 clinical cases (450 CAP and 450 normal), achieving accuracy, sensitivity, specificity, and F1-Score of 0.946, 0.942, 0.949, and 0.944, respectively. In a similar approach, Mei et al. [22] proposed a joint AI algorithm to combine chest CT findings with clinical data (symptoms, exposure history, and laboratory testing) to diagnose COVID-19 from non-COVID patients using a dataset of 905 cases. The joint model achieved high discriminative performance with 0.92 AUC, 84.3% sensitivity, and 82.8% specificity, outperforming a senior radiologist who achieved 0.84 AUC, 74.6% sensitivity, and 93.8% specificity. The drawback of combined systems is the availability of clinical information especially when a large number of suspected patients are waiting to be diagnosed. Furthermore, the proposed studies do not show the infection location in the lung which can be useful for the medical personnels for longitudinal monitoring of the patients.

The aforementioned machine learning solutions with CT imaging were limited to only COVID-19 detection. However, COVID-19 pneumonia screening is important for evaluating the status of the patient and treatment. In particular, COVID-19 related infection localization and the segmentation of pneumonia lesions is a crucial task for accurate diagnosis and follow-up of pneumonia patients. Zhou et al. [23] proposed a lesion detection system that can quantify COVID-19 infection regions from the chest CT scans. Three independent two-dimensional (2D) U-Nets are used for x-y, y-z, and x-z views of CT scan, where for each model, five adjacent slices are used as an input, while the network outputs infection prediction mask for the middle slice. The three intermediate binary predictions are aggregated by a simple sum up, with a threshold value of 2 to detect infection pixels. Moreover, to alleviate the data scarcity for annotated infection masks, a dynamic model was developed for data augmentation by simulating the progression of infection regions using multiple CT scan readings from the same patient. With the augmented data, the proposed system showed a performance of 78.3% DSC and 77.6% sensitivity. Besides, deep learning has a high potential to automate the lesion detection task but requires a large set of high-quality annotations that are difficult to collect during the current pandemic. Learning from noisy training labels that are easier to generate has the potential to alleviate this problem. Wang et al. [24] introduced a novel framework to learn from noisy COVID-19 infection masks. They first proposed a new Dice loss metric, which integrates Dice loss and Mean Absolute Error (MAE). Then a novel COVID-19 pneumonia lesion segmentation network (COPLE-Net) was developed that can segment COVID-19 infected regions with various scales and appearances. Moreover, an adaptive self-ensembling training strategy was proposed, which outperforms standard deep learning training strategies in scenarios of learning from noisy segmentation labels. The proposed framework achieved promising segmentation results with Dice, Relative Volume Error (RVE), and 95 percentile of Hausdorff Distance (HD-95) of 80.72%, 15.96%, and 17.12 ± 29.35 mm, respectively. HD-95 is the distance between the segmentation results and the ground truth in 3D space. Wang et al. [25] proposed a weakly supervised deep learning framework for COVID-19 classification and infection localization. A three-stage framework was introduced, where first the lung regions were segmented using a pre-trained 2D U-Net model slice by slice; then a proposed deep convolutional neural network (DeCoVNet) was used to classify the entire 3D lung volume to COVID-19 or non-COVID. Finally, COVID-19 lesions were localized by integrating the activation regions in the classification network obtained by gradient class activation map (Grad-CAM), with activation maps from the lung segmentation model obtained by a 3D connected components method (3DCC). Therefore, the proposed algorithm does not require annotated infection masks in the training phase, as no dedicated segmentation model is used for infection localization, while ground-truth masks were provided by a professional radiologist for test set only to evaluate the network performance. The introduced algorithm was trained and evaluated on a dataset of only 313 COVID-19, and 229 non-COVID cases achieving classification results of 0.959 AUC value, and 0.976 precision-recall (PR)-AUC. However, a poor infection segmentation performance was reported with a 68.5% of hit rate (HR). Authors in [26] proposed a system that carried out lung and lesion segmentation for CT images using DRUNET, which provides a DSC value of 95.9% for lung segmentation. On the other hand, lesion segmentation scored a mean DSC (mDSC) of 58.7% for DeepLabv3 [27] using 4,695 CT slice images were used for lung and lesion segmentation. The performance indicators for the previous studies in lesion segmentation are lower than the lung segmentation, which can be improved further. A large annotated dataset is required to increase the performance as only 201 scans were used in the results reported in [23]. Similar problems and computationally expensive techniques were reported in research articles [25] and [26].

For severity quantification, several studies recommended that using deep learning can help in the quantification of COVID-19 lung opacification. Moreover, it can eliminate the subjectivity in the initial assessment for COVID-19 patients. Chaganti et al. [28] presented a method that automatically segments and quantifies abnormal CT patterns in COVID-19 patients. The proposed system utilized 9749 chest CT volume and segmented lesions, lungs, and lobe areas and used four matrices for severity quantification: percentage of opacity, percentage of high opacity, lung severity score, and lung high opacity score. Despite the good performance, no clear evaluation metric for segmentation network models was presented. Another work classified the severity into four classes (mild, moderate, severe, and critical) [16]. Lung and lesion segmentation were carried out using the UNet model via commercial tools with a median DSC of 0.85% for both models. Shen et al. [29] created a system that considers computer and radiologist evaluations to determine the COVID-19 patient severity. The computer approach consisted of four phases: segmentation of the lung and lobes, segmentation of the pulmonary vessels, filter out pulmonary vessels from the lung region, and detection of infection. The lesion segmentation was done using thresholds and adaptive region growing. The work showed that the Pearson correlation between computer and radiologist evaluation was ranged from 0.7679 to 0.8373, which was carried out using only 44 patients. Pu et al. [30] created an automated system to quantify COVID-19 severity and progression using chest CT images. 120 patients were used to train and evaluate two U-Net models for lung and vessel segmentation. The proposed system achieved 95% and 81% DSC for lung and lesion segmentation, respectively. It is notable that the model failed to deal with pneumonic regions that are very small and near the vessels. Besides, the work used small datasets for training and testing, a total of 192 CT volume were used in this work.

Although most of the reviewed studies showed good performance for both lung and infection segmentation tasks, they mainly used conventional U-Net architecture or other techniques that are based on image processing. However, recently, different variants of U-Net architecture and other encoder-decoder (E-D) CNN, such as feature pyramid network (FPN), with residual, dense blocks, or inception blocks have shown state-of-the-art segmentation results in various applications. Therefore, there is still room to investigate the capability of those architectures for lung detection and COVID-19 infection localization tasks. Besides, several studies used a small number of patients and CT images to train, test, and validate the proposed systems. Table 1 summarizes the results of segmentation and classification obtained by the recent studies in the literature, and the table highlights the dataset size and the main networks used in each study.

Table 1. A quantitative comparison between the state of the art for classification and segmentation tasks.

| | DATASET | | NETWORK | | | | RESULTS | | 3D VISUALIZATION |
|---|---|---|---|---|---|---|---|---|---|
| REF | NON-COVID SCANS (SUBJECTS) | COVID-19 SCANS (SUBJECTS) | COVID-19 CLASSIFICATION | LUNG SEGMENTATION | LESION SEGMENTATION | IDENTIFICATION | LUNG SEGMENTATION [DSC] | INFECTION SEGMENTATION [DSC] | |
| [16] | - | 300 (126) | - | - | UNet | - | - | 84.81% | |
| [31] | 498 (498) | 1420 (704) | DCN | UNet | UNet | 89.62% Sens. 98.04% Spec. | 99.11% | 83.51% | |
| [32] | - | 110 (60) | - | - | UNet | - | - | 82.00% | |
| [33] | 1092(628) | 1094(960) | ResNet34 + Size-balanced Sampling | VB-Net | VB-Net | 86.9% Sens 90.1% Spec | 98.00% | 92.00% | |
| [20] | 1,695 (1,695) | 1,029 (922) | 3D Densnet121 | 3D AH-Net | – | 85.30% Sens. 90.10% Spec. | 95.00% | – | |

| [34] | 6,814 (5,941) | 4,542 (3,084) | ResNet152 | UNet | - | 96.36% Sens. 80.09% Spec. | 92.55% | - | |
| [23] | _ | 201 (140) | - | - | 2.5D UNet | - | - | 78.30% | ✓ |
| [24] | - | 558 (558) | - | - | COPLE-Net | - | - | 80.72% | ✓ |

*1.2. Motivation*

Although the above studies have demonstrated some promising results by using chest CT for the diagnosis of COVID-19, there is room for improvement particularly in lesion segmentation and severity detection. Several works addressed lung and lesion segmentation, as shown in the previous section, which can help physicians to diagnose COVID-19 accurately and to assess the treatment response. The performance of the lesion segmentation models is still low compared to lung segmentation. This work aims to propose a system to identify and classify the severity of COVID-19 patients into four levels: mild, moderate, severe, and critical infection. Besides, the work investigates different deep learning methods for detecting COVID-19 infected slices from CT volume. For segmentation, U-Net [35] and feature pyramid network (FPN) [36] models were investigated with different encoders to achieve the best performance for lung and lesion segmentation. ResNet18 [37], ResNet50, ResNet152 [37], DenseNet121, DenseNet161, and DenseNet201 [38] were used as the backbone encoder for the segmentation models. Additionally, a reliable method was proposed to identify COVID-19 slices from the prediction maps generated by infection segmentation models. Besides, COVID-19 infection is quantified by computing the percentage of infected lung pixels on the segmented lung CT slices. Finally, a 3D volumetric visualization is developed to show the overall infected area in the lungs. This work uses several datasets from 1,139 patients (51,027 CT slices) for training and validation and thus the system dealt with different images from different devices with varying image quality levels.

The rest of the paper is organized as follows: Section II describes the used methodology adopted for the study. The experimental setup and evaluation metrics are presented in Section III. Section IV presents the results and performs an extensive set of comparative evaluations among the networks employed and we discuss and analyze the results. Finally, the conclusions are drawn in Section V.

## 2 Methodology

The proposed system consists of three main stages as shown in Figure 1, where the segmentation of lung from CT images is the first step of our proposed system. Transfer learning was used on encoder layers with ImageNet weights to train the segmentation networks. The input CT volumes are evaluated slice-by-slice. First, a binary lung mask is generated for input CT slice using the 1st E-D CNN. Next, the lung is segmented using the generated mask and fed to the 2nd E-D CNN, which identifies the infected lung regions. The generated infection mask is used to detect COVID-19 slices from the normal slices. Besides, COVID-19 pneumonia lesions are localized using the generated lung and infection masks. The output lesion model is used to identify COVID -19 slices from normal slices. Furthermore, the infection percentage in the lung is found for the patient in order to classify the severity of the given volume into four classes based on the infection percentage of the lung. Finally, a visualization tool is used to visualize the infection areas within the patients' lungs. This section first presents the datasets used in this study. Then we shall introduce the pre-processing techniques applied to these datasets, different machine learning models investigated, and the quantification technique of COVID-19 infection.

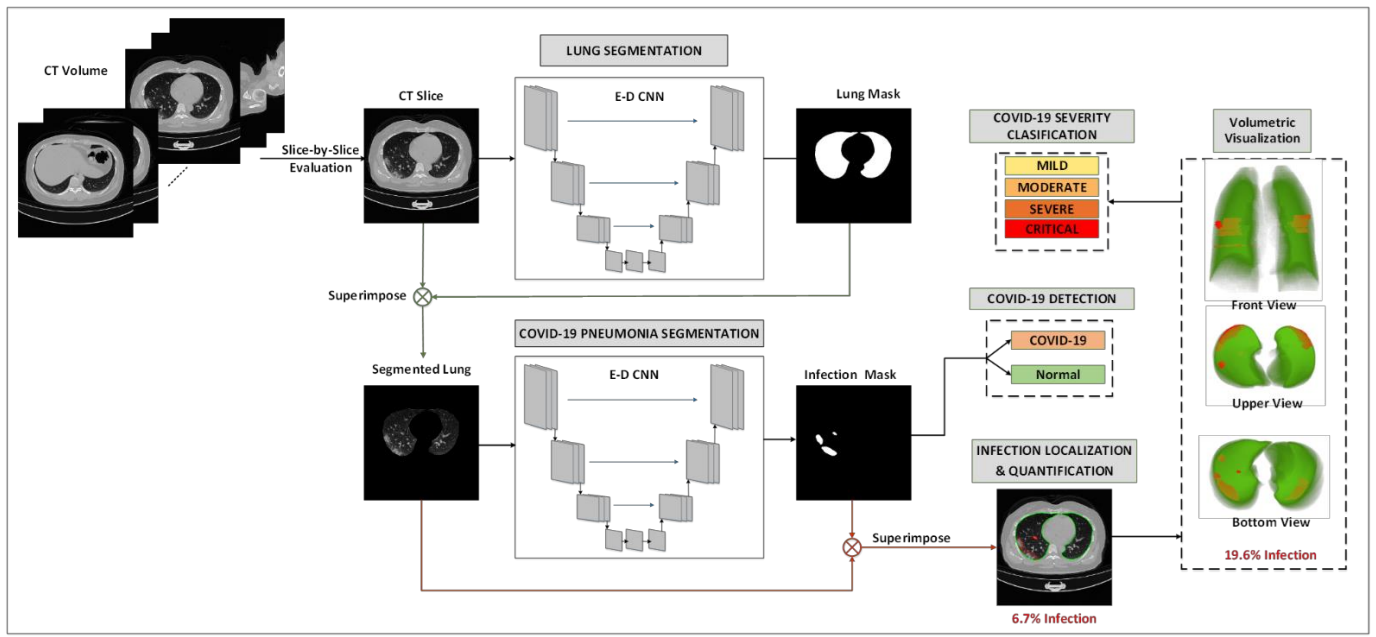

*Figure 1. Schematic representation of the pipeline of the proposed COVID-19 recognition system*

## 2.1 CT Datasets

To train and evaluate the proposed system, four public datasets from different sources were used in this work (Table 2). A total of 1,139 patients and 51,027 CT slices were used in this work. The description of the used datasets is below:

The first dataset [39] consists of CT volumes from 20 patients including 3520 CT images with ground truth lung masks and lesion masks. These images were labeled by two radiologists and verified by an experienced radiologist as mentioned in the dataset description. The second dataset called "COVID-19 CT segmentation dataset" [40], the dataset is based on volumetric CTs from Radiopaedia. It includes 9 patients with 829 slices along with their corresponding ground truth lung masks, which are created by expert radiologists. Another dataset was found on the Kaggle platform, it consists of 267 CT slices with their corresponding ground truth lung masks [41], the images are non-COVID cases as they were collected in 2017. Additionally, MosMedData [42] was used for external validation. The dataset was obtained between 1$^{st}$ of March, 2020 and 25$^{th}$ of April, 2020, and provided by the medical hospitals in Moscow, Russia. It includes 1,110 patients where 50 patients have been annotated by the experts to show infection areas in 784 CT slices. The dataset consists of normal (254 patients) and COVID-19 cases (856 patients), the COVID-19 cases are split into 4 classes CT1 (affected lung percentage 25% or below, 684 images), CT2 (from 25% to 50%, 125 patients), CT3 (from 50% to 75%, 45 patients), and CT4 (75% and above, 2 patients).

Table 2. Summary of the datasets used in this work

| Dataset Name | Task | # of Patients | # Images used in this study | Lung Mask | Lesion Mask |
|---|---|---|---|---|---|
| COVID-19 CT Lung and Infection Segmentation Dataset | Lung segmentation, lesion segmentation, & COVID-19 detection | COVID-19: 20 | 3520 | ✓ | ✓ |
| COVID-19 CT segmentation dataset | Lung segmentation | COVID-19: 9 | 829 | ✓ | |
| Finding and Measuring Lungs in CT Data (Kaggle) | Lung segmentation | Not available | 267 | ✓ | |
| MosMedData* | External Validation | 1110 | 46,411 | | ✓ |

*The dataset creators provided 50 cases with ground truth lesion masks

## 2.2 Pre-processing

Three out of four datasets provided the CT images in Neuroimaging Informatics Technology Initiative (NIfTI) format. However, different window levels (in Hounsfield units (HU)) were specified for the different datasets. This creates the following issue: the image features are not consistent across different datasets. Therefore, all NIfTI files were converted into Portable Network Graphics (PNG) format images, and the image intensity values have been normalized and mapped to pixel values in the range of 0-255, then intensity interval has been changed for each dataset to create consistent image content. Finally, all images were resized to 256×256 for the segmentation tasks. Figure 2 shows the sample images in each dataset.

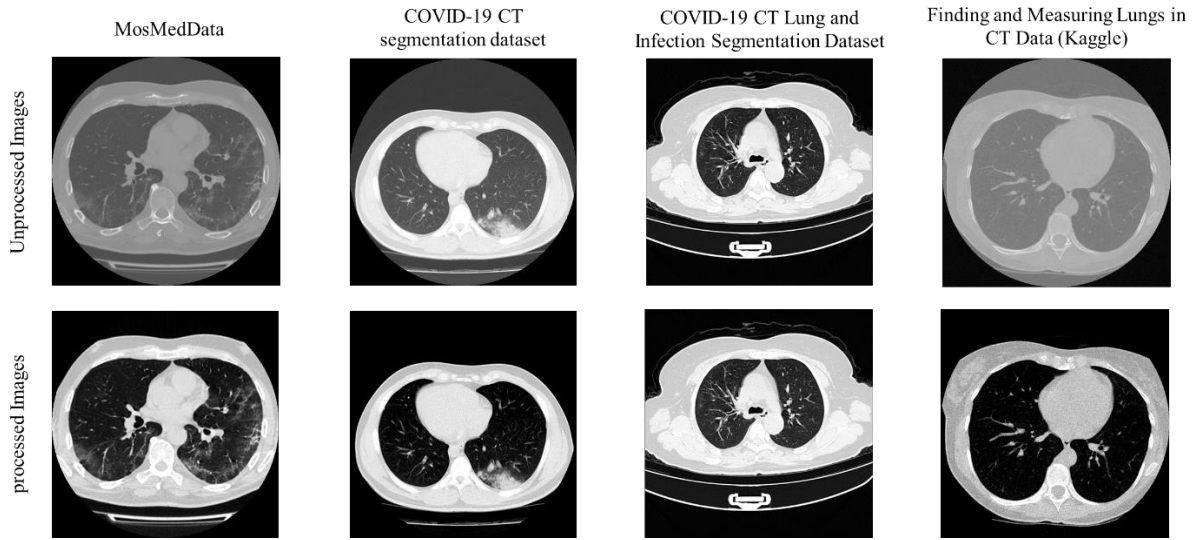

*Figure 2. Sample of processed and unprocessed CT images used in this work*

## 2.3   Network models for lung and COVID-19 infection segmentation and classification

Firstly, a deep learning model is developed to generate a lung mask for the input CT slice. The segmented lung is then fed to another deep learning model to identify the infection regions within the segmented CT image. The produced infection mask is used to detect COVID slices. Furthermore, the COVID-19 infection lesion is quantified by computing the percentage of infected lung pixels and visualized on the 3D volumetric model.

Lung parenchyma and COVID-19 infections segmentation were performed on CT slices using the state-of-the-art deep Encoder-Decoder Convolutional Neural Networks (E-D CNNs), U-Net, and FPN, with different backbone (encoder) models using the variants of DenseNet and ResNet. Several variants of the two backbone models were considered starting from shallow to deep structures: ResNet18, ResNet50, ResNet152, DenseNet121, DenseNet161, and DenseNet201.

The utilized encoder-decoder architecture provides a powerful segmentation model that captures the context in the contracting path and enables precise localization by the expanding path. For U-Net architecture, 1×1 convolution is utilized to map the output from the last decoding block to two-channel feature maps, where a pixel-wise SoftMax activation function is applied to map each pixel into a binary class of background or lung for Lung parenchyma segmentation task, and background or lesion for infection segmentation task. While FPN employs the encoder and decoder structure as a pyramidal hierarchy where a prediction mask is made on each spatial level of the decoder path. In the final step, predicted feature maps are up-sampled to the same size, concatenated, convolved with a 3×3 convolutional kernel, and SoftMax activation is applied to generate the final prediction mask. Transfer learning was utilized on the encoder side of the segmentation networks by initializing the convolutional layers with ImageNet weights.

*Segmentation Loss function*

The cross-entropy (CE) loss is used as the cost function for the segmentation networks:

$$CE = -\frac{1}{K}\sum_k \sum_c y_k \log(p(x_k)) \qquad (1)$$

where $x_k$ denotes the $k^{th}$ pixel in the predicted segmentation mask, $p(x_k)$ denotes its SoftMax probability, $y_k$ is a binary random variable getting 1 if $y_k = c$, otherwise 0, and $c$ denotes the class category, i.e., $c \in \{background, lung\}$ for the lung segmentation task, and $c \in \{background, lesion\}$ for the infection segmentation.

## 2.4 The proposed approach for COVID-19 detection and severity classification

The detection of COVID-19 is performed based on the prediction maps generated by the lesion segmentation networks. Accordingly, a CT slice is classified as COVID-19 positive if at least one pixel is predicted as COVID-19 infection, i.e., $p(x_k) > 0.5$, otherwise, the image is considered normal. The severity of the

COVID-19 patient is classified into four classes based on lung parenchyma percentage in the patients' lungs: mild, moderate, severe, and critical infection. The Percentage of Infection (PI) is calculated as the infected areas (sum of white pixels) over the lung area for one CT slice. For the entire volume, the average of all slices is considered as the patient severity percentage. Based on the percentage, the patient is classified into four classes. Figure 3 demonstrates the process of calculation of Percentage of Infection (PI) on one CT slice.

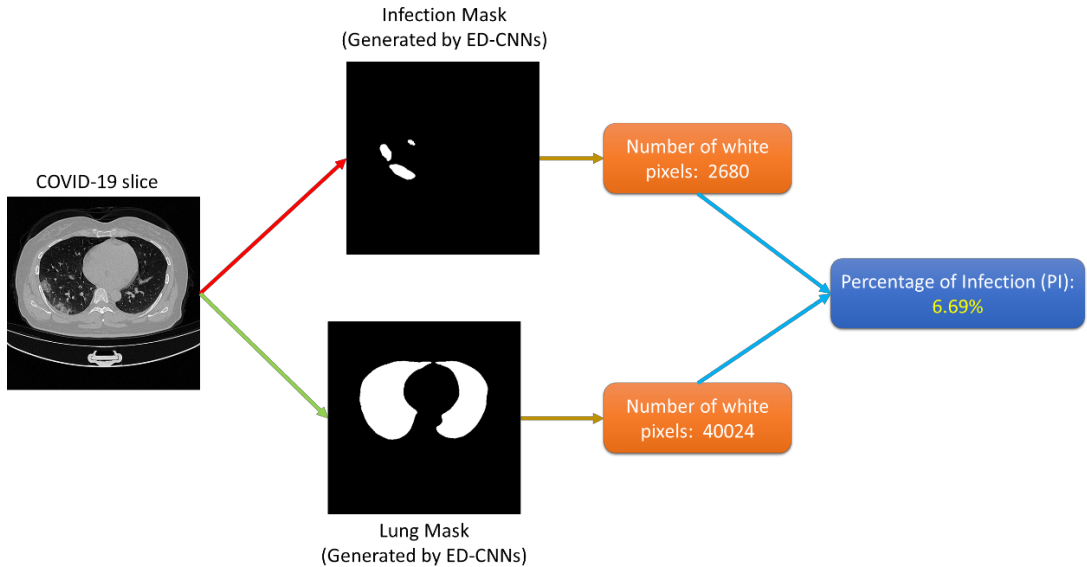

*Figure 3. Proposed approach to calculate the infection percentage for CT image*

## 3 Experimental Setup

Classification and segmentation models were implemented using PyTorch library with Python 3.7 on Intel® Xeon® CPU E5-2697 v4 @2.30GHz and 64 GB RAM, with an 8-GB NVIDIA GeForce GTX 1080 GPU card. Segmentation models were trained using Adam optimizer with learning rate, $\alpha = 10^{-3}$, momentum updates, $\beta_1 = 0.9$ and $\beta_2 = 0.999$, mini-batch size of 4 images with 50 backpropagation epochs as shown in Table 3. Early stopping criterion was used as follows: when no improvement in validation loss is seen during the 10 epochs, training is stopped abruptly. Table 3 presents the training and hyper-parameters for the lung and infection segmentation models.

Table 3. Details of lung/ lesion segmentation models training parameters

| TRAINING PARAMETERS | LUNG SEGMENTATION MODEL | INFECTION SEGMENTATION MODEL |
|---|---|---|
| Batch size | 4 | 4 |
| Learning rate | 0.001 | 0.001 |
| Number of folds | 5 | 10 |
| Learning rate drop factor | 0.2 | 0.2 |
| Max epochs | 50 | 50 |
| Epochs patience | 5 | 5 |
| Epochs stopping criteria | 10 | 10 |
| Optimizer | Adam | Adam |
| Function Loss | NLLLoss | NLLLoss |

## 3.1 Data preparation and Augmentation

Lung Segmentation networks were trained using 5-fold cross-validation (CV), with 80% train and 20% test (unseen) folds, where 20% of training data was used as a validation set to avoid overfitting. For infection segmentation, instead of 5 fold cross-validation, 10 fold cross-validation was used. Class imbalance in the dataset impacts the performance of the deep learning models. Thus, data augmentation was used to balance the size of each class in lung and lesion segmentation datasets to ensure every possible aspect of avoiding data overfitting [43]. This step is crucial for the training phase to reduce the associated error from the lung segmentation task, which might propagate to the subsequent lesion segmentation task [44]. We performed data augmentation by applying rotations of 90, -90, 180 degrees for CT images and ground truth masks. Table 4 summarizes the number of images per class used for training, validation, and testing at each fold. Independent training and evaluation were provided for both the networks, where original CT slices were used as input to the lung segmentation models, and lung segmented CT slices were used as inputs to the lesion segmentation network, where infection masks were used as ground-truth. Besides, a combined evaluation was provided using the best lung segmentation and infection segmentation models to evaluate the overall performance of the proposed cascaded system.

Table 4. Number of CT mages per class and per fold before and after data augmentation

| Task | Class | # of Samples | Training Samples | Augmented Training Samples | Validation Samples | Test Samples |
|---|---|---|---|---|---|---|
| Lung Segmentation | CT with corresponding lung masks | 4616 | 2955 | 11820 | 923 | 738 |
| Lesion Segmentation | CT with corresponding lesion masks | 3520 | 2253 | 9012 | 563 | 704 |

## 3.2 Evaluation Criteria

Quantitative evaluations for the proposed approach are performed for lung segmentation, infection segmentation, and COVID-19 detection tasks. The segmentation tasks are evaluated on the pixel-level, where the foreground (lung or infected region) was considered as the positive class, and background as the negative class. For the COVID-19 detection task, the performance was computed per CT sample, where slices with COVID-19 infection were considered as the positive class and normal slices were considered as the negative class.

The performance of detection and segmentation tasks was assessed using different evaluation metrics with 95% confidence intervals (CIs). Accordingly, the CI for each evaluation metric was computed as follows:

$$r = z\sqrt{metric(1-metric)/N} \qquad (2)$$

where, $N$ is the number of test samples, and $z$ is the level of significance that is 1.96 for 95% CI. All values were computed over the overall confusion matrix that accumulates all test fold results of the 5-fold or 10-fold cross-validation in respective experiments.

The performance of the lung and lesion segmentation networks were evaluated using three evaluation metrics which are accuracy, Intersection over Union (IoU), and Dice Similarity Coefficient (DSC):

$$Accuracy = \frac{TP + TN}{TP + TN + FP + FN} \qquad (3)$$

where $accuracy$ is the ratio of the correctly classified pixels among the image pixels. *TP, TN, FP, FN* represent the true positive, true negative, false positive, and false negative, respectively.

$$Intersection\ over\ Union\ (IoU) = \frac{TP}{TP + FP + FN} \tag{4}$$

$$Dice\ Similarity\ Coefficient\ (DSC) = \frac{2TP}{2TP + FP + FN} \tag{5}$$

where, both $IoU$ and $DSC$ are statistical measures of spatial overlap between the binary ground-truth segmentation mask and the predicted segmentation mask, whereas the main difference is that DSC considers double weight for $TP$ pixels (true lung/lesion predictions) compared to IoU.

Five evaluation metrics were considered for the COVID-19 detection scheme: accuracy, sensitivity, precision, F1-score, and specificity.

$$Precision = \frac{TP}{TP + FP} \tag{6}$$

where $precision$ is the rate of correctly classified positive class CT samples among all the samples classified as positive samples.

$$Sensitivity = \frac{TP}{TP + FN} \tag{7}$$

where $sensitivity$ is the rate of correctly predicted positive samples in the positive class samples,

$$F1 = 2\frac{Precision \times Sensitivity}{Precision + Sensitivity} \tag{8}$$

where $F1$ is the harmonic average of precision and sensitivity.

$$Specificity = \frac{TN}{TN + FP} \tag{9}$$

where $specificity$ is the ratio of accurately predicted negative class samples to all negative class samples.

# 4 Results and Discussion

This section describes the results of the lung and lesion segmentation, COVID-19 detection and severity classification along with 3D lung modeling to visualize lung infections.

## 4.1 *Lung Segmentation*

The results of 5-fold cross-validation are tabulated in Table 5. For each model, it was observed that three encoders: DenseNet, 121, 161, and 201 are the top-performing ones for lung segmentation. However, it is visible that the FPN network with different encoders did not improve the results compared with different U-Net architectures, which is a standard network for segmentation tasks. U-Net with DenseNet, 121, 161, and 201 encoders showed the best DSC performance for lung segmentation. DenseNet 121 is the best-performing network for lung segmentation with IoU and DSC of 95.35% and 97.11% receptively. The outputs of the top three networks compared with the ground truth are shown in Figure 4. It can be observed that the segmentation of U-Net with DenseNet, 121, 161, and 201 is highly consistent with the ground truth. An interesting observation is the ability of the three networks for creating the segmentation mask for the small lung region. This is considered a challenging task for the deep learners as shown in rows, 2 and 3 in Figure 4, where it was shown that the network can generate a mask for the small lung slices. Although the lungs can be severely affected by COVID-19 lesions, the trained model successfully segmented the lung boundaries as shown in Figure 4. This reflects the robustness of the model proposed in this study for lung segmentation. Authors in [22, 45] discarded the small lung area (less than 20% of the body part) slices during the pre-processing phase. But this work included such images in the training and testing sets.

Table 5. Results of 5- fold cross-validation of Lung Segmentation

| NETWORK | ACCURACY (%) | IoU (%) | DSC (%) |
|---|---|---|---|
| UNet | 99.70 ± 0.16 | 95.04 ± 0.63 | 96.61 ± 0.52 |
| ResNet18 UNet | 99.70 ± 0.16 | 95.01 ± 0.63 | 96.84 ± 0.5 |
| ResNet50 UNet | 99.70 ± 0.16 | 95.03 ± 0.63 | 96.8 ± 0.51 |
| ResNet152 UNet | 99.70 ± 0.16 | 94.95 ± 0.63 | 96.69 ± 0.52 |
| **DenseNet 121 UNet** | **99.70 ± 0.16** | **95.35 ± 0.61** | **97.11 ± 0.48** |
| **DenseNet 161 UNet** | **99.69 ± 0.16** | **95.10 ± 0.62** | **97.19 ± 0.48** |
| **DenseNet 201 UNet** | **94.88 ± 0.64** | **94.88 ± 0.64** | **97.00 ± 0.49** |
| ResNet18 FPN | 99.65 ± 0.17 | 93.6 ± 0.71 | 95.76 ± 0.58 |
| ResNet50 FPN | 99.65 ± 0.17 | 93.39 ± 0.72 | 95.52 ± 0.6 |
| ResNet152 FPN | 99.66 ± 0.17 | 93.92 ± 0.69 | 96.00 ± 0.57 |
| DenseNet 121 FPN | 99.67 ± 0.16 | 94.53 ± 0.66 | 96.55 ± 0.53 |
| DenseNet 161 FPN | 99.66 ± 0.17 | 94.05 ± 0.68 | 96.11 ± 0.56 |
| DenseNet t201 FPN | 99.67 ± 0.17 | 94.35 ± 0.67 | 96.39 ± 0.54 |

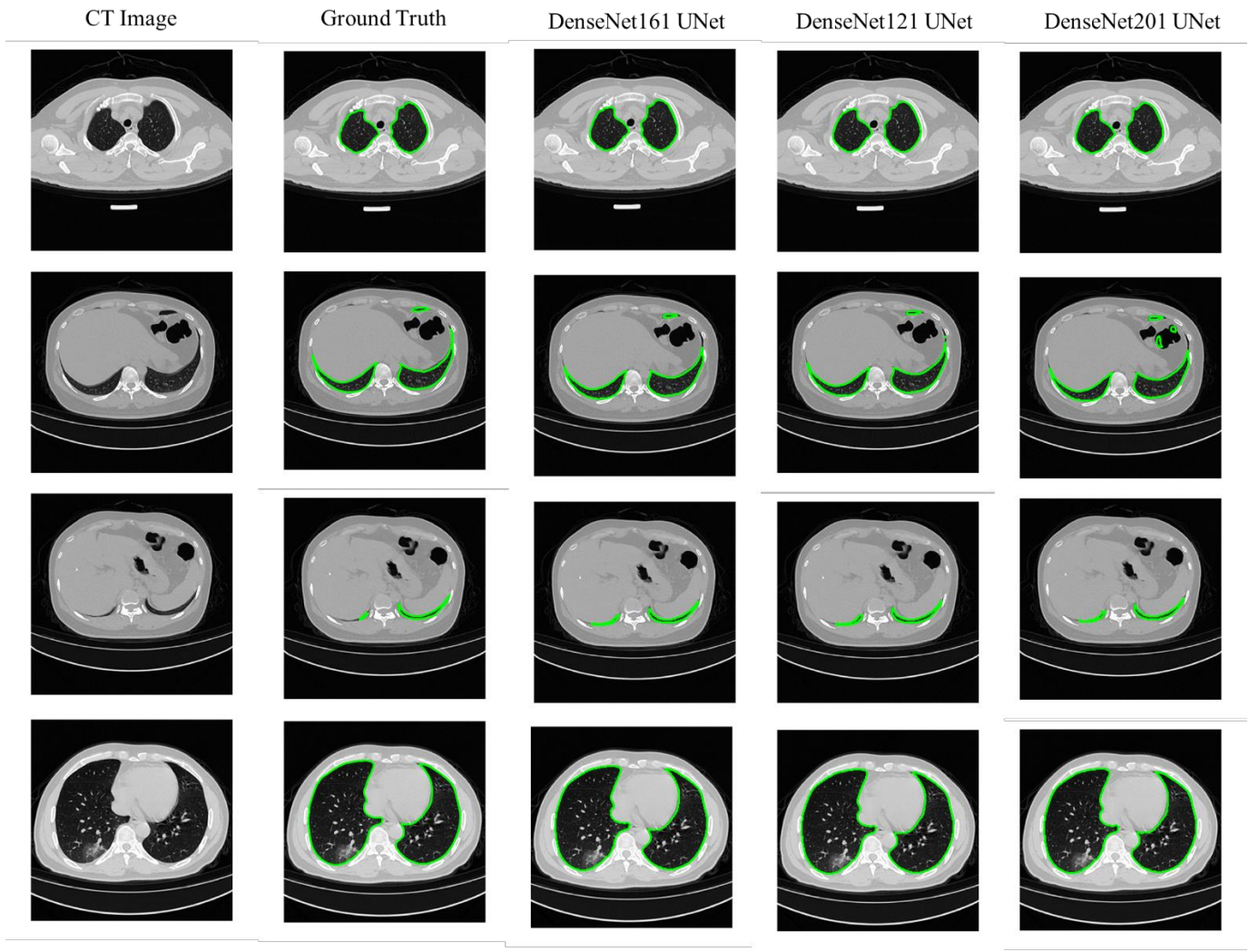

*Figure 4. CT image (1st row), ground truth (2nd row), and the segmentation masks of the top three networks (rows 3-5).*

## 4.2 Lesion Segmentation

Lesion segmentation can assist medical doctors to diagnose better the infection in the lung. The segmentation performances of the different networks are presented in Table 6. The results indicate that the FPN network performs better than the UNet in general. DenseNet201 FPN achieved the best segmentation performance with IoU, and DSC of 91.85% and 94.13%, respectively. The second and third best networks were FPN models too but the results are very close with insignificant differences. Figure 6 shows the ability of the top three networks to segment the infected regions even from small lung areas (Figure 6).

Table 6. Results of 10- fold cross-validation of Lesion Segmentation

| NETWORK | ACCURACY (%) | IoU (%) | DSC (%) |
|---|---|---|---|
| UNet | 99.82 ± 0.18 | 90.2 ± 0.72 | 92.52 ± 0.6 |
| ResNet18 UNet | 99.82 ± 0.18 | 90.69 ± 0.72 | 92.97 ± 0.58 |
| ResNet50 UNet | 98.56 ± 0.18 | 89.09 ± 0.72 | 91.44 ± 0.58 |
| ResNet152 UNet | 99.8 ± 0.18 | 88.41 ± 0.72 | 90.80 ± 0.59 |
| DenseNet121 UNet | 99.81 ± 0.18 | 90.58 ± 0.7 | 92.88 ± 0.55 |
| DenseNet161 UNet | 99.82 ± 0.18 | 90.86 ± 0.71 | 93.07 ± 0.55 |
| DenseNet201 UNet | 99.82 ± 0.73 | 91.13 ± 0.73 | 93.36 ± 0.56 |
| **ResNet18 FPN** | **99.81 ± 0.2** | **91.45 ± 0.81** | **93.80 ± 0.67** |
| **ResNet50 FPN** | **99.81 ± 0.2** | **91.46 ± 0.82** | **93.82 ± 0.68** |
| ResNet152 FPN | 99.8 ± 0.19 | 90.66 ± 0.79 | 93.05 ± 0.65 |
| DenseNet121 FPN | 99.68 ± 0.19 | 89.09 ± 0.75 | 91.02 ± 0.6 |
| DenseNet161 FPN | 99.81 ± 0.19 | 91.11 ± 0.78 | 93.45 ± 0.64 |
| **DenseNet201 FPN** | **99.81 ± 0.19** | **91.85 ± 0.76** | **94.13 ± 0.62** |

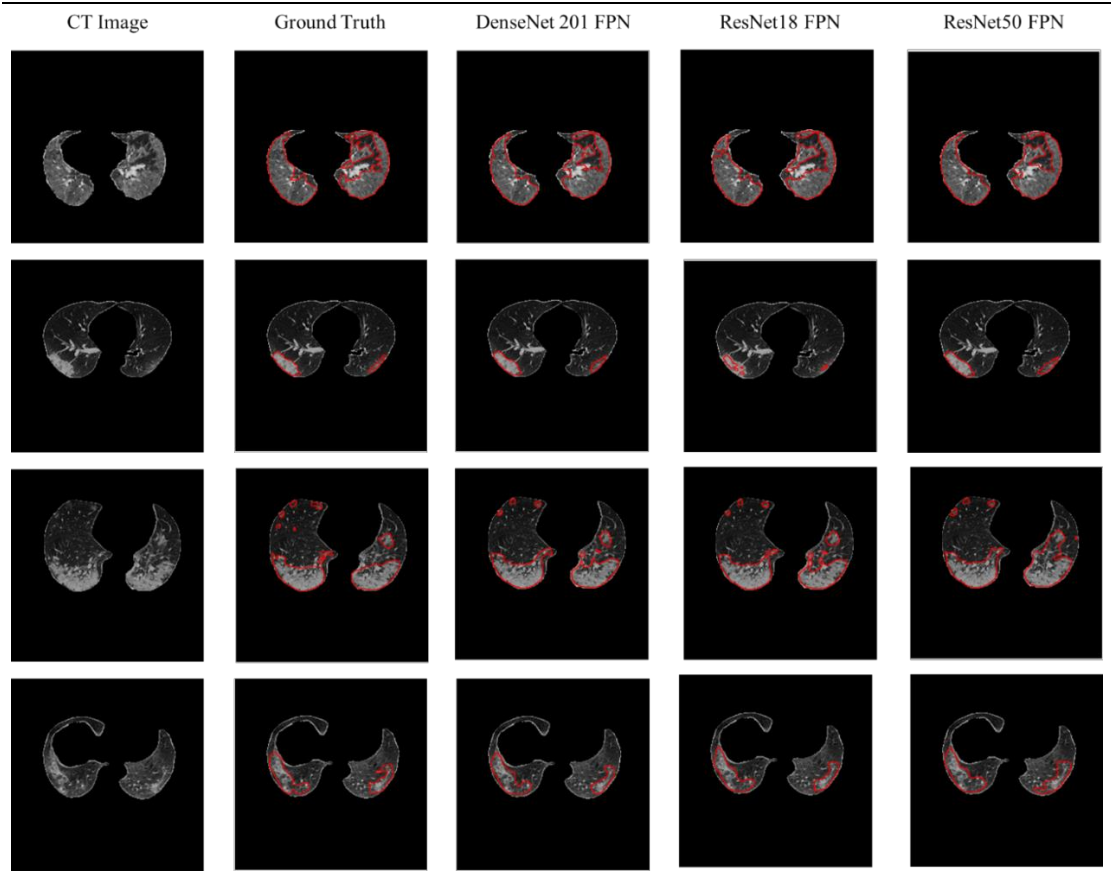

Figure 5. CT image (1$^{st}$ row), ground truth (2$^{nd}$ row), and the lesion segmentation of the top three networks (rows 3-5).

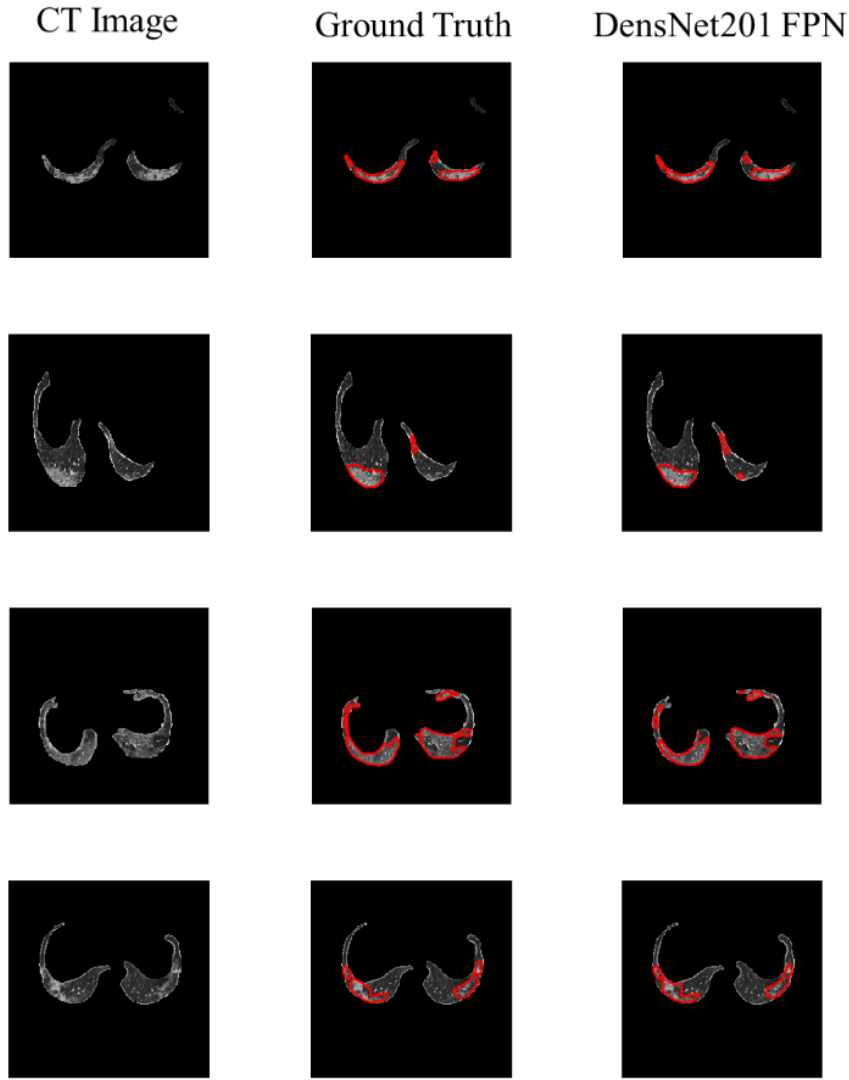

*Figure 6. CT image (1st row), ground truth (2nd row), and the lesion segmentation of the best network (row 3)*

## 4.3  COVID-19 detection

The performance of lesion segmentation networks from the CT lung images is presented in Table 7. Since missing any COVID-19 positive case is critical, sensitivity is the primary metric that we consider in detection. All the networks achieved high sensitivity values (>99%), where both U-Net and FPN networks with DenseNet201 as the backbone achieved the best performance with 99.64% sensitivity, which indicates that the proposed approach can achieve a high level of robustness. Moreover, the FPN model with DenseNet201 as backbone achieved the specificity of 98.72%, indicating a significantly low false alarm rate.

Table 7. The detection performance of the lesion segmentation networks

| NETWORK | ACCURACY | PRECISION | SENSITIVITY | F1-SCORE | SPECIFICITY |
|---|---|---|---|---|---|
| UNet | 95.23 ± 0.7 | 84.92 ± 1.18 | 99.35 ± 0.27 | 91.57 ± 0.92 | 93.77 ± 0.8 |
| ResNet18 UNet | 94.6 ± 0.75 | 84.03 ± 1.21 | 99.78 ± 0.15 | 91.32 ± 0.93 | 92.46 ± 0.87 |
| ResNet50 UNet | 95.66 ± 0.67 | 85.91 ± 1.15 | 99.69 ± 0.18 | 92.29 ± 0.88 | 94.24 ± 0.77 |
| ResNet152 UNet | 95.68 ± 0.67 | 86.06 ± 1.14 | 99.56 ± 0.22 | 92.32 ± 0.88 | 94.31 ± 0.77 |
| DenseNet121 UNet | 95.86 ± 0.66 | 86.59 ± 1.13 | 99.53 ± 0.23 | 92.61 ± 0.86 | 94.57 ± 0.75 |
| DenseNet161 UNet | 95.61 ± 0.68 | 85.9 ± 1.15 | 99.26 ± 0.28 | 92.1 ± 0.89 | 94.35 ± 0.76 |
| DenseNet201 UNet | 95.93 ± 0.65 | 86.67 ± 1.12 | 99.64 ± 0.2 | 92.7 ± 0.86 | 94.63 ± 0.74 |
| ResNet18 FPN | 95.74 ± 0.67 | 86.25 ± 1.14 | 99.4 ± 0.26 | 92.36 ± 0.88 | 94.46 ± 0.76 |
| ResNet50 FPN | 95.77 ± 0.66 | 86.29 ± 1.14 | 99.45 ± 0.24 | 92.4 ± 0.88 | 94.49 ± 0.75 |
| ResNet152 FPN | 98.44 ± 0.41 | 95.07 ± 0.72 | 99.48 ± 0.24 | 97.23 ± 0.54 | 98.04 ± 0.46 |
| DenseNet121 FPN | 97.16 ± 0.55 | 91.67 ± 0.91 | 99 ± 0.33 | 95.19 ± 0.71 | 96.43 ± 0.61 |
| DenseNet161 FPN | 97.05 ± 0.56 | 90.62 ± 0.96 | 98.91 ± 0.34 | 94.58 ± 0.75 | 96.39 ± 0.62 |
| **DenseNet201 FPN** | **98.96 ± 0.34** | **96.47 ± 0.61** | **99.64 ± 0.2** | **98.03 ± 0.46** | **98.72 ± 0.37** |

## *4.4  Severity classification using MosMedData Dataset*

1,110 patients were provided in MosMedData Dataset, which was used to test the performance of the proposed severity classification system. Lung and lesion masks were generated using the best-performing networks obtained from previous sections. Figure 7 shows three examples of predicted masks by both models: the best lung segmentation network (DenseNet 161 UNet) and the best lesion segmentation network (DenseNet201 FPN) on an entirely independent dataset. It can be seen that the cascaded networks were able to detect the lung borders very accurately and also performed well in detecting the main COVID-19 infection regions.

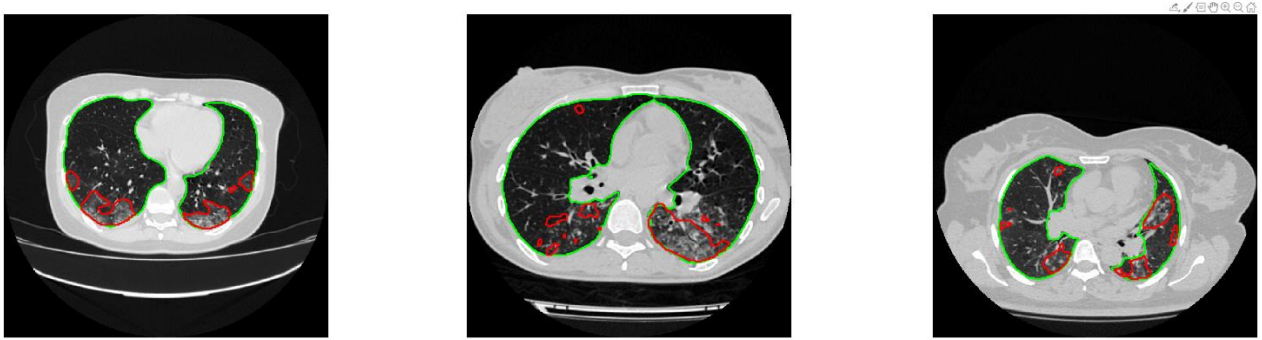

*Figure 7. Detection of lung and lesion for external validation.*

The infection percentage has been calculated for each CT volume, where each volume is classified as healthy (CT0), or with mild (CT1), moderate (CT2), severe (CT3), or critical (CT4) COVID-19 infection using the criteria mentioned in the MosMedData Dataset, however, quantified using our infection percentage quantification method. It should be noted that the ground truth of CT0-CT4 classification was provided in the dataset, which was done by visually inspecting the CT slices by professional radiologists. The confusion matrix for the classification of 1,110 patients is shown in Figure 8 and the quantitative evaluation is summarised in Table 8.

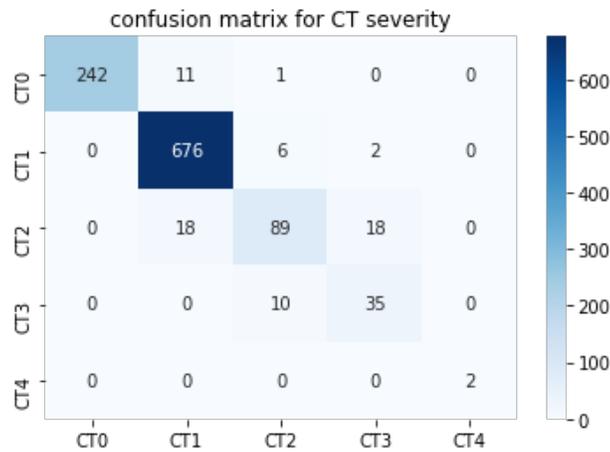

Figure 8. Confusion matrix for CT severity classification of the CT volumes of MosMedData Dataset.

Table 8. Performance matrices for Severity classification

| CLASS | INFECTION (%) | ACCURACY (%) | PRECISION (%) | SENSITIVITY (%) | F1-SCORE (%) | SPECIFICITY (%) |
|---|---|---|---|---|---|---|
| CT0 | Healthy | 98.92 ± 1.27 | 100 ± 0 | 95.28 ± 2.61 | 97.58 ± 1.89 | 100 ± 0 |
| CT1 | 0 < x < 25 | 96.67 ± 1.35 | 95.89 ± 1.49 | 98.83 ± 0.81 | 97.34 ± 1.21 | 93.19 ± 1.89 |
| CT2 | 25 < x < 50 | 95.23 ± 3.74 | 83.96 ± 6.43 | 71.2 ± 7.94 | 77.06 ± 7.37 | 98.27 ± 2.28 |
| CT3 | 50 < x < 75 | 97.3 ± 4.74 | 63.64 ± 14.06 | 77.78 ± 12.15 | 70 ± 13.39 | 98.12 ± 3.97 |
| CT4 | 75 < x < 100 | 100 ± 0 | 100 ± 0 | 100 ± 0 | 100 ± 0 | 100 ± 0 |
| Average | | 97.05 ± 1 | 94.18 ± 1.38 | 94.05 ± 1.39 | 94 ± 1.4 | 95.53 ± 1.22 |

From the confusion matrix, it can be observed that the system can reliably classify severe (CT4) volumes with 100% accuracy. Moreover, the majority of normal cases, CT0, were classified correctly, only 12 cases were misclassified, whereas 8 out of the 12 studies showed an infection percentage of 2-4%. However, those patients may have abnormalities other than COVID-19 in the lung. In fact, in the data description of MosMedData Dataset [42], it was said that no viral pneumonia is shown in these cases. However, other types of pneumonia may exist. Figure 8 shows that all COVID-19 cases were detected as CT1, CT2, CT3, and CT4; none of the COVID-19 cases were predicated as normal (CT0). In other words, the system can distinguish COVID-19 patients from healthy cases on an independent test set very reliably. Thus, the severity classification performance matches the results obtained in the detection section. Furthermore, the proposed system showed lower sensitivity values for moderate (CT2) and severe (CT3) compared to CT0, CT1, and CT4, this can be related to how the dataset was labeled by radiologists. The dataset was labeled by two radiologists using a visual semi-quantitative approach, such an approach can lead to weak labeling [46].

### 4.5 3D Modelling of Lung Volume with Infection Visualization

A 3D model of the lung with infection segmentation is generated for each patient using the output of lung and lesion segmentation networks. The proposed tool can assist the medical doctors to assess better the infection and to evaluate its severity. Figure 9 shows 3D lung models from different views while the COVID-19 infection is presented with the red color saturation.

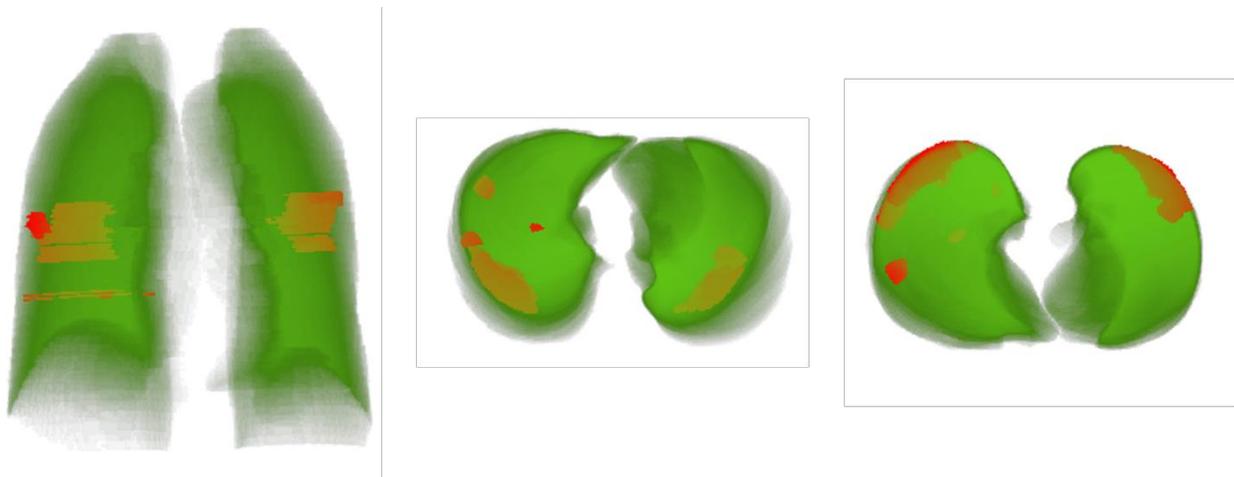

*Figure 9: The proposed 3D lung models from different views whilst the infection area is marked in red.*

## 5 Conclusion

In this paper, we proposed a systematic approach for COVID-19 detection, lung, and lesion segmentation, and patients' severity grading from the CT images. To find the best performing deep learning models, we have investigated several state-of-the-art segmentation networks. The proposed approach with the cascaded models achieved elegant performance levels in segmentation, classification, infection quantification, and 3D visualization. The main conclusions of this study can be summarized as follows:

- For lung segmentation, DenseNet 161 U-Net outperformed the FPN network with different encoders. On the other hand, FPN with DenseNet201 encoder performed the best in lesion segmentation with a DSC value of 94.13%.
- The proposed lesion segmentation pipeline can generate better lung and lesion masks for small lung areas. Such images and masks are typically excluded by many studies in the literature.
- The right choice of encoders can significantly boost the performance of segmentation models. This study demonstrated that the DenseNet family outperforms ResNet in lung segmentation.
- The proposed approach with the FPN DenseNet201 encoder model achieved the highest sensitivity of 99.64% in COVID-19 detection performance.
- The system was able to classify the severity for COVID-19 patients based on Percentage of Infection (PI) by considering the output of lung and lesion segmentation networks and was able to discriminate between different severity levels of COVID-19 infection over a dataset of 1,110 subjects with sensitivity values of 98.3%, 71.2%, 77.8% and 100% for mild, moderate, severe and critical infections, respectively.
- In summary, computer-aided detection and quantification is an accurate, easy, and feasible method to diagnose COVID-19 cases.

## Funding

Qatar University COVID19 Emergency Response Grant (QUERG-CENG-2020-1) provided the support for the work and the claims made herein are solely the responsibility of the authors.

## Declaration of Competing Interest

The authors report no declarations of interest.

# References

[1] "Weekly epidemiological update on COVID-19 , 15 December 2020 " in "Emergency Situational Updates," World Health Organization2020, Available: https://www.who.int/publications/m/item/weekly-epidemiological-update---15-december-2020.

[2] P. Craw and W. Balachandran, "Isothermal nucleic acid amplification technologies for point-of-care diagnostics: a critical review," *Lab on a Chip,* 10.1039/C2LC40100B vol. 12, no. 14, pp. 2469-2486, 2012.

[3] V. M. Corman *et al.*, "Detection of 2019 novel coronavirus (2019-nCoV) by real-time RT-PCR," (in eng), *Euro surveillance : bulletin Europeen sur les maladies transmissibles = European communicable disease bulletin,* vol. 25, no. 3, p. 2000045, 2020.

[4] P. Kakodkar, N. Kaka, and M. N. Baig, "A Comprehensive Literature Review on the Clinical Presentation, and Management of the Pandemic Coronavirus Disease 2019 (COVID-19)," (in eng), *Cureus,* vol. 12, no. 4, pp. e7560-e7560, 2020.

[5] G. D. Rubin *et al.*, "The Role of Chest Imaging in Patient Management during the COVID-19 Pandemic: A Multinational Consensus Statement from the Fleischner Society," *Radiology,* vol. 296, no. 1, pp. 172-180, 2020/07/01 2020.

[6] S. Salehi, A. Abedi, S. Balakrishnan, and A. Gholamrezanezhad, "Coronavirus Disease 2019 (COVID-19): A Systematic Review of Imaging Findings in 919 Patients," *American Journal of Roentgenology,* vol. 215, no. 1, pp. 87-93, 2020/07/01 2020.

[7] Y. Fang *et al.*, "Sensitivity of Chest CT for COVID-19: Comparison to RT-PCR," *Radiology,* vol. 296, no. 2, pp. E115-E117, 2020/08/01 2020.

[8] T. Ai *et al.*, "Correlation of chest CT and RT-PCR testing in coronavirus disease 2019 (COVID-19) in China: a report of 1014 cases," *Radiology,* p. 200642, 2020.

[9] H. Shi *et al.*, "Radiological findings from 81 patients with COVID-19 pneumonia in Wuhan, China: a descriptive study," *The Lancet Infectious Diseases,* vol. 20, no. 4, pp. 425-434, 2020.

[10] A. Esteva *et al.*, "Dermatologist-level classification of skin cancer with deep neural networks," vol. 542, no. 7639, pp. 115-118, 2017.

[11] H. Dong, G. Yang, F. Liu, Y. Mo, and Y. Guo, "Automatic brain tumor detection and segmentation using U-Net based fully convolutional networks," in *annual conference on medical image understanding and analysis*, 2017, pp. 506-517: Springer.

[12] L. Shen, L. R. Margolies, J. H. Rothstein, E. Fluder, R. McBride, and W. J. S. r. Sieh, "Deep learning to improve breast cancer detection on screening mammography," vol. 9, no. 1, pp. 1-12, 2019.

[13] D. Ardila *et al.*, "End-to-end lung cancer screening with three-dimensional deep learning on low-dose chest computed tomography," vol. 25, no. 6, pp. 954-961, 2019.

[14] A. Tahir *et al.*, "Coronavirus: Comparing COVID-19, SARS and MERS in the eyes of AI," *arXiv preprint arXiv:1706.05587,* 2020.

[15] X. Xu *et al.*, "A Deep Learning System to Screen Novel Coronavirus Disease 2019 Pneumonia," *Engineering,* 2020/06/27/ 2020.

[16] L. Huang *et al.*, "Serial Quantitative Chest CT Assessment of COVID-19: Deep-Learning Approach," *Radiology: Cardiothoracic Imaging,* vol. 2, no. 2, p. e200075, 2020/04/01 2020.

[17] D. P. Fan *et al.*, "Inf-Net: Automatic COVID-19 Lung Infection Segmentation From CT Images," *IEEE Transactions on Medical Imaging,* vol. 39, no. 8, pp. 2626-2637, 2020.

[18] H. Kang *et al.*, "Diagnosis of Coronavirus Disease 2019 (COVID-19) With Structured Latent Multi-View Representation Learning," *IEEE Transactions on Medical Imaging,* vol. 39, no. 8, pp. 2606-2614, 2020.

[19] S. Wang *et al.*, "A Fully Automatic Deep Learning System for COVID-19 Diagnostic and Prognostic Analysis," *European Respiratory Journal,* p. 2000775, 2020.

[20] S. A. Harmon *et al.*, "Artificial intelligence for the detection of COVID-19 pneumonia on chest CT using multinational datasets," vol. 11, no. 1, pp. 1-7, 2020.

[21] Q. Wang, D. Yang, Z. Li, X. Zhang, and C. J. I. A. Liu, "Deep Regression via Multi-Channel Multi-Modal Learning for Pneumonia Screening," vol. 8, pp. 78530-78541, 2020.

[22] X. Mei *et al.*, "Artificial intelligence–enabled rapid diagnosis of patients with COVID-19," pp. 1-5, 2020.

[23] L. Zhou *et al.*, "A rapid, accurate and machine-agnostic segmentation and quantification method for CT-based covid-19 diagnosis," *IEEE Transactions on Medical Imaging,* vol. 39, no. 8, pp. 2638-2652, 2020.

[24] G. Wang *et al.*, "A noise-robust framework for automatic segmentation of covid-19 pneumonia lesions from ct images," *IEEE Transactions on Medical Imaging,* vol. 39, no. 8, pp. 2653-2663, 2020.

[25] X. Wang *et al.*, "A Weakly-supervised Framework for COVID-19 Classification and Lesion Localization from Chest CT," *IEEE Transactions on Medical Imaging,* 2020.

[26] K. Zhang *et al.*, "Clinically Applicable AI System for Accurate Diagnosis, Quantitative Measurements, and Prognosis of COVID-19 Pneumonia Using Computed Tomography," *Cell,* vol. 181, no. 6, pp. 1423-1433.e11, 2020.

[27] L.-C. Chen, G. Papandreou, F. Schroff, and H. Adam, "Rethinking atrous convolution for semantic image segmentation," *arXiv preprint arXiv:1706.05587,* 2017.

[28] S. Chaganti *et al.*, "Automated Quantification of CT Patterns Associated with COVID-19 from Chest CT," *Radiology: Artificial Intelligence,* vol. 2, no. 4, p. e200048, 2020/07/01 2020.

[29] C. Shen *et al.*, "Quantitative computed tomography analysis for stratifying the severity of Coronavirus Disease 2019," *Journal of Pharmaceutical Analysis,* vol. 10, no. 2, pp. 123-129, 2020/04/01/ 2020.

[30] J. Pu *et al.*, "Automated quantification of COVID-19 severity and progression using chest CT images," *European Radiology,* vol. 31, no. 1, pp. 436-446, 2021/01/01 2021.

[31] K. Gao *et al.*, "Dual-branch combination network (DCN): Towards accurate diagnosis and lesion segmentation of COVID-19 using CT images," *Medical Image Analysis,* vol. 67, p. 101836, 2021/01/01/ 2021.

[32] X. Chen, L. Yao, and Y. Zhang, "Residual Attention U-Net for Automated Multi-Class Segmentation of COVID-19 Chest CT Images," *arXiv preprint arXiv:2004.05645,* 2020.

[33] X. Ouyang *et al.*, "Dual-Sampling Attention Network for Diagnosis of COVID-19 From Community Acquired Pneumonia," *IEEE Transactions on Medical Imaging,* vol. 39, no. 8, pp. 2595-2605, 2020.

[34] C. Jin *et al.*, "Development and evaluation of an artificial intelligence system for COVID-19 diagnosis," *Nature Communications,* vol. 11, no. 1, p. 5088, 2020/10/09 2020.

[35] O. Ronneberger, P. Fischer, and T. Brox, "U-net: Convolutional networks for biomedical image segmentation," in *International Conference on Medical image computing and computer-assisted intervention*, 2015, pp. 234-241: Springer.

[36] T.-Y. Lin, P. Dollár, R. Girshick, K. He, B. Hariharan, and S. Belongie, "Feature pyramid networks for object detection," in *Proceedings of the IEEE conference on computer vision and pattern recognition*, 2017, pp. 2117-2125.

[37] K. He, X. Zhang, S. Ren, and J. Sun, "Deep residual learning for image recognition," in *Proceedings of the IEEE conference on computer vision and pattern recognition*, 2016, pp. 770-778.

[38] G. Huang, Z. Liu, L. Van Der Maaten, and K. Q. Weinberger, "Densely connected convolutional networks," in *Proceedings of the IEEE conference on computer vision and pattern recognition*, 2017, pp. 4700-4708.

[39] M. Jun *et al.* COVID-19 CT Lung and Infection Segmentation Dataset [Online]. Available: https://zenodo.org/record/3757476

[40] COVID-19 CT segmentation dataset [Online]. Available: http://medicalsegmentation.com/covid19/

[41] K. S. Mader. Finding and Measuring Lungs in CT Data [Online]. Available: https://www.kaggle.com/kmader/finding-lungs-in-ct-data/data

[42] S. P. Morozov *et al.*, "MosMedData: Chest CT Scans With COVID-19 Related Findings Dataset," *arXiv preprint arXiv:2005.06465,* 2020.

[43] M. Krebbs, *Deep Learning With Python* (Data Sciences). CreateSpace Independent Publishing Platform, 2018.

[44] J. Wang *et al.*, "Prior-Attention Residual Learning for More Discriminative COVID-19 Screening in CT Images," *IEEE Transactions on Medical Imaging,* vol. 39, no. 8, pp. 2572-2583, 2020.

[45] M. Rahimzadeh, A. Attar, and S. M. Sakhaei, "A Fully Automated Deep Learning-based Network For Detecting COVID-19 from a New And Large Lung CT Scan Dataset," *medRxiv,* p. 2020.06.08.20121541, 2020.

[46] M. Goncharov *et al.*, "CT-based COVID-19 Triage: Deep Multitask Learning Improves Joint Identification and Severity Quantification," *arXiv preprint arXiv:2006.01441,* 2020.